\documentclass[12pt]{article}
\begin{document}

\title{On complex adaptive systems and terrorism}
\author{\ E. Ahmed$^{1}$, A. S. Elgazzar$^{2}$ and A. S. Hegazi$^{1}$ \\
$^{1.}$Mathematics Department, Faculty of Science\\
35516 Mansoura, Egypt\\
$^{2.}$Mathematics Department, Faculty of Education\\
45111 El-Arish, Egypt}
\maketitle
\date{}

\begin{abstract}
Complex adaptive systems (CAS) are ubiquitous in nature. They are
basic in social sciences. An overview of CAS is given with
emphasize on the occurrence of bad side effects to seemingly
"wise" decisions. Hence application to terrorism is given. Some
conclusions on how to deal with this phenomena are proposed.
\end{abstract}

\section{Introduction}

Terrorism is an important phenomena that deserves to be studied
using all possible approaches. Here we use complex adaptive
systems (CAS) [Boccara 2004] to study it. We use some aspects of
CAS and present some conclusions about terrorism. One of our main
conclusions is that total eradication of terrorism is highly
unlikely. It is more feasible to localize it. In sec. 2  CAS are
reviewed. Terrorism is then defined. Then we apply CAS to
terrorism using game theory with mistakes [Sato Crutchfield 2002,
Ahmed et al 2003]. Some comments are given on the work of Galam et
al [Galam 2003, Galam and Mauger 1996] who uses percolation theory
[Aharony and Stauffer 1992] to study terrorism. In sec. 3 some
general conclusions on CAS are presented then used to give several
proposals on handling terrorism.

\section{Complex adaptive systems and terrorism}

\textbf{Definition (1):} A complex adaptive system consists of
inhomogeneous, interacting adaptive agents.\\
\\
\textbf{Definition (2):} An emergent property of a CAS is a
property of the system as a whole which does not exist at the
individual elements (agents) level.\\

Typical examples are the brain, the immune system [Matzinger 2002,
Segel and Cohen 2001], the economy, social systems, ecology
[Edelstein-Keshet 1988], insects swarm, etc…

The existence of emergent properties implies that to understand a
complex system one has to study the system as a whole and not to
decompose it into its constituents. This totalistic approach is
against the standard reductionist one, which tries to decompose
any system to its constituents and hopes that by understanding the
elements one can understand the whole system.\\
\\
\textbf{Why should we study complex adaptive systems?}

Most of the real systems are CAS. Moreover they have intrinsic
unpredictability which causes some "seemingly wise" decisions to
have harmful side effects. Therefore we should try to understand
CAS to try to minimize such side effects.\\
\\
\textbf{How to model a CAS?}

The standard approaches are
\begin{enumerate}
    \item Ordinary differential equations
(ODE), difference equations and partial differential equations
(PDE).
    \item Cellular automata (CA) [Ilachinski 2001].
    \item Evolutionary game theory [Hofbauer and Sigmund 1998].
    \item Agent based models.
    \item Networks [Watts and Strogatz 1998] etc..
\end{enumerate}
Some of these approaches are included in [Boccara  2004].
\\
\\
\\
\textbf{Application to terrorism:}

First we define what we mean by terrorism:\\
\\
\textbf{Definition (3):} Terrorism is the attack on unarmed
civilians. It
is obvious that terrorism is a CAS.\\

Recently [Sato and Crutchfield 2002] have studied the dynamics of
learning in multi-agent systems, where the agents use
reinforcement learning. Their work was extended to evolutionary
game theory by [Ahmed et al 2003]. Here we apply Sato Crutchfield
idea to spiteful replicator dynamics so we propose the following
form
\begin{equation}\label{1}
\frac{dx_{i}}{dt}=x_{i}\sum_{j=1}^{n}x_{j}\left( \Pi _{ij}-\Pi
_{ji}\right) +\gamma _{i}x_{i}\sum_{j=1}^{n}x_{j}\ln \left(
\frac{x_{j}}{x_{i}}\right) ,
\end{equation}
where $\gamma _{i},i=1,2,...,n$ are nonnegative constants
measuring the average rate of mistakes done by the player adopting
strategy $i$.
    An immediate  result  from Eq. (1) is that $x_{i}\neq 0,i=1,2,...,n$. This means that eradication of (even wrong) ideas is extremely hard and that it is more practical to contain them. An obvious example is the idea of Nazism which has been beaten militarily more than 50 years ago yet it has not been eradicated.

    Another interesting idea proposed recently by Galam [Galam 2003] is to use percolation theory [Stauffer and Aharony 1992] to understand terrorism and propose strategies to fight it. Consider a matrix of passive sympathizers who are not involved into any acts of terror yet they share the terrorists' motives but not acts. Terrorism will then be like a percolating phenomena in this matrix. If fraction of support $p$ exceeds a given threshold say $p>p_{c}$, then terror will propagate but if $p<p_{c}$, then it will be contained. The bright idea of Galam is in realizing that to contain terror  one may either affect the connectivity of  the passive supporters $q$ which is extremely difficult or by reducing the dimension of the social space $d$ which equals 2(space) + number of motives (called flags) which the passive sympathizers share with the terrorists. He proposed the formula
 \begin{equation}\label{2}
 p_{c}=a\left[ (q-1)(d-1)\right] ^{-b},a=1.287,b=0.616,
\end{equation}
where $q=16, d=10$. We expect that $d=6$ since there are four main
flags which can be identified as: Occupation of two countries,
feeling that religion is being attacked and supporting some
repressive states. This makes $p_{c}\approx 10\%$ which agrees
with Galam estimates.

\section{Conclusions}
Before a decision (concerning a CAS) is made the following points
should be taken into considerations:
\begin{enumerate}
    \item[(i)] CAS should be studied as a
whole hence reductionist point of view may not be reliable in some
cases.
    \item[(ii)] CAS are open with nonlinear local interactions hence:
    \begin{enumerate}
        \item Long range prediction is highly unlikely [Strogatz 2000,
Holmgren 1996].
        \item When studying a CAS take into consideration the effects of
its perturbation on related systems. This is also relevant to the
case of natural disasters where an earthquake at a city can cause
a widespread power failure at other cities.
        \item Expect side effects to any "WISE" decision.
        \item Mathematical and computer models may be helpful in
reducing such side effects.
    \end{enumerate}
    \item[(iii)] Optimization in CAS should be
multi-objective and not single objective [Collette and Siarry
2003].
    \item[(iv)] CAS are very difficult to control.
Interference at highly connected sites may be a useful approach
[Dogoretsev and Mendez 2004]. The interlinked nature of CAS
elements complicates both the unpredictability and controllability
problems. It also plays an important role in innovations spread.
    \item[(v)] Memory effects should not be
neglected in CAS. Also memory games have been studied [Smale 1980,
Ahmed and Hegazi 2000].
\end{enumerate}

    Now how can these ideas help in proposing strategies in the war against terror?\\
\\
\\
\\
\textbf{We propose the following:}
\begin{enumerate}
    \item Terrorism should be studied as a whole. Security solutions
alone cannot be successful since it does not affect $p_{c}$. In
fact excessive force may increase $d$. This agrees with many
observations. Changing d is possible through combined political,
economic and social in addition to security actions.
    \item Total eradication of terror is almost impossible.
Sato-Crutchfield game shows that every possible strategy will be
used. Thus containment and not eradication is the feasible goal.
    \item Memory effects should be taken into considerations
hence do not expect quick solutions.
    \item Target interference to affect terror networks may
be more effective than an all out one.
    \item Expect side effects and some failures.
    \item CAS are open and distributed [Kaneko
1993] systems hence take into consideration the effect of  a
decision on related people.
\end{enumerate}

\section*{References}

\begin{enumerate}
\item[ ] Ahmed E. and Hegazi A.S.(2000)"On Discrete Dynamical
Systems Associated to Games", Adv. Complex. Sys. 2, 423.

\item[ ] Ahmed E., Hegazi A.S., Elgazzar A.S.(2003),
Sato-Crutchfield formulation for some Evolutionary Games, Int. J.
Mod. Phys. C 14, 963.

\item[ ] Boccara N. (2004), Modeling complex systems, Springer
Publ, Berlin.

\item[ ] Collette Y. and Siarry P. (2003), "Multiobjective
Optimization" Springer.

\item[ ] Dogoretsev S.N. and Mendes J.F (2004), "The shortest path
to complex networks", Cond-mat 0404593.

\item[ ] Edelstein-Keshet L. (1988), Introduction to mathematical
biology,Random House N.Y.

\item[ ] Galam S. (2003), "Global Physics. From percolation to
terrorism", Physica A 330, 139.

\item[ ] Galam S. and Alain Mauger, Phys. Rev. E 53 (1996) 2177S.

\item[ ] Holmgren R. (1996), "A first course in  discrete
dynamical systems" Springer.

\item[ ] Hofbauer J. and Sigmund K. (1998), "Evolutionary games
and population dynamics" Cambridge Univ.Press U.K.

\item[ ] Ilachinski A.(2001),''Cellular automata'',World
Scientific Publ, Singapore.

\item[ ] Kaneko K. (1993), "Theory and application of coupled map
lattices" Wiley Pub.

\item[ ] Matzinger P. (2002), "The danger model" Science 296,12,
pp. 301.

\item[ ] Sato Y. and Crutchfield J. P. (2002), Coupled replicator
equations for the dynamics of learning in multi-agent systems,
Phys. Rev.E.67, 40.

\item[ ] Segel L.A. and Cohen I.R.(eds.)(2001), Design principles
for the immune system and other distributed autonomous systems,
Oxford Univ. Press U.K.

\item[ ] Smale, S.(1980). The prisoners' dilemma and dynamical
systems associated to games. Econometrica 48, 1617-1634.

\item[ ] Smith J.B.(2003), ''Complex systems'', CS 0303020.

\item[ ] Stauffer D. and Aharony A. (1992), "Introduction to
percolation theory", Taylor and Francis, U.K.

\item[ ] Strogatz S. (2001), "Nonlinear dynamics and chaos",
Perseus Books Group.

\item[ ] Watts D.J. and Strogatz S.H. (1998), "Collective dynamics
of small world network", Nature 393, 440.
\end{enumerate}

\end{document}